\documentstyle[12pt,epsf]{article}
\renewcommand{\bar}[1]{\overline{#1}}

\begin{document}

\begin{flushright}
USM-TH-77,
hep-ph/9906424, Version to appear in PLB
\end{flushright}

\bigskip\bigskip
\centerline{\large \bf Flavor and Spin Structure of
$\Lambda$-Baryon at Large x}

\vspace{22pt}

\centerline{\bf Bo-Qiang Ma$^{a}$, Ivan Schmidt$^{b}$, and
Jian-Jun Yang$^{b,c}$}

\vspace{8pt}
{\centerline {$^{a}$Department of Physics, Peking University,
Beijing 100871, China,\footnote{Mailing address}}}

{\centerline {CCAST (World Laboratory), P.O.~Box 8730, Beijing
100080, China,}}

{\centerline {and Institute of High Energy Physics, Academia Sinica,
P.~O.~Box 918(4),}}

{\centerline {Beijing 100039, China}}


{\centerline {$^{b}$Departamento de F\'\i sica, Universidad
T\'ecnica Federico Santa Mar\'\i a,}}

{\centerline {Casilla 110-V, 
Valpara\'\i so, Chile}}


{\centerline {$^{c}$Department of Physics, Nanjing Normal
University,}}

{\centerline {Nanjing 210097, China}}



\vspace{10pt}
\begin{center} {\large \bf Abstract}

\end{center}

It is shown that a perturbative QCD (pQCD) based analysis 
and the SU(6)
quark-diquark model give significant different predictions
concerning the flavor and spin structure for the quark
distributions of the $\Lambda$-baryon near $x=1$. Detailed
predictions for the ratios $u(x)/s(x)$ of unpolarized quark
distributions, $\Delta s(x)/s(x)$ of valence strange quark, and
$\Delta u(x)/u(x)$ of valence up and down quarks of the $\Lambda$
are given from the quark-diquark model and from a pQCD based
model. It is found that the up and down quarks are positively
polarized at large $x$, even though their net spin contributions
to the $\Lambda$ might be zero or negative. The significant
difference for $u(x)/s(x)$ between the two different 
approaches are predicted. 
The prediction of positively
polarized up and down quarks inside the $\Lambda$ at large $x$
has been supported
by the available data of $\Lambda$-polarization in $Z$ decays and
also by the most recent HERMES result of spin transfer to the $\Lambda$
in deep elastic scattering of polarized lepton on the nucleon target.

\vfill

\centerline{PACS numbers: 14.20.Jn, 12.38.Bx, 12.39.Ki, 13.60.Hb}

\vfill
\newpage


Although it is well established knowledge that hadrons are
composite systems of quarks and gluons, the detailed quark
structure of hadrons remains a domain with many unknowns, and
there have been many unexpected surprises with respect to naive
theoretical considerations. The sea content of the nucleons has
received extensive investigations concerning its spin structure
\cite{SpinR}, strange content \cite{Bro88,Bro96}, flavor asymmetry
\cite{Kum97}, and isospin symmetry breaking \cite{Isospin}, and it
is commonly taken for granted that our understanding of the
valence quark structure of the nucleons is more clear. However,
even in this last case the situation remains doubtful, reflected
in the recent investigations concerning the flavor and spin
structure of the valence quarks for the nucleon near $x=1$. For
example, there are different predictions concerning the ratio
$d(x)/u(x)$ at $x \to 1$ from a perturbative QCD (pQCD) based analysis
\cite{Far75,Bro95} and the SU(6) quark-diquark model
\cite{DQM,Ma96,Mel96}, and there are different predictions
concerning the value of $F_2^n(x)/F_2^p(x)$ at large $x$, which
has been taken to be $1/4$ as in the quark-diquark model in most
parameterizations of quark distributions. A recent analysis
\cite{Yang99} of experimental data from several processes suggests
that $F_2^n(x)/F_2^p(x) \to 3/7$ as $x \to 1$, in favor of the
pQCD based prediction. The spin structure of the valence quarks is also
found to be different near $x=1$ in these models, and predictions
have been made concerning the non-dominant valence down ($d$)
quark, so that $\Delta d(x)/d(x)=-1/3$ in the quark-diquark model
\cite{Ma96,Mel96}, a result which is different from the pQCD based
prediction $\Delta q(x)/q(x)=1$ for either $u$ and $d$
\cite{Bro95}. At the moment, there is still no clear data in order
to check these different predictions, although the available
measurements \cite{SMC96} for the polarized $d$ quark
distributions seem to be negative at large $x$, slightly in favor
of the quark-diquark model prediction.

In this letter, we show that the same mechanisms that produce the
different flavor and spin structure for the quark distributions of
the nucleon, give also significant different predictions
concerning the flavor and spin structure for the quark
distributions of the $\Lambda$-baryon near $x=1$, thus providing
tests of different approaches. 
We also show
that the non-dominant up ($u$) and down ($d$) quarks of the
$\Lambda$ should be positively polarized at large $x$, 
even though their net spin contributions
to the $\Lambda$ might be zero or negative. In fact, it was found by
Burkardt and Jaffe \cite{Bur93} that the $u$ and $d$ quarks should
be negatively polarized from SU(3) symmetry. Recently, it was also
pointed out by Soffer and one of us \cite{Ma99} that the flavor
and spin content of the $\Lambda$ can be used to test different
predictions concerning the spin structure of the nucleon and the
quark-antiquark asymmetry of the nucleon sea. Thus it is clear
that the quark structure of $\Lambda$ is a frontier with rich
physics and deserves further attention both theoretically and
experimentally.

We now look into the details of the flavor and spin structure for
the valence quarks of the $\Lambda$. We start our analysis in the
SU(6) quark-diquark model. We know that exact SU(6) symmetry in
the SU(6) quark model predicts $u(x)=2d(x)$ for the proton and
this gives the prediction $F_2^n(x)/F_2^p(x) \geq 2/3$ for all
$x$. This result was ruled out by the experimental observation
that $F_2^n(x)/F_2^p(x) < 1/2$ for $x \to 1$, where the valence
quark contributions are dominant. The SU(6) quark-diquark model
\cite{DQM} introduces a breaking to the exact SU(6) symmetry by
the mass difference between the scalar and vector diquarks and
predicts $d(x)/u(x) \to 0$ at $x \to 1$, leading to a ratio
$F_2^n(x)/F_2^p(x) \to 1/4$ which could fit the data and has been
accepted in most parameterizations of quark distributions for the
nucleon. In this work we analyze the valence quark distributions
of the $\Lambda$ by extending the SU(6) quark-spectator-diquark
model \cite{Ma96}, which can be considered as a revised version of
the original SU(6) quark-diquark models \cite{DQM}, from the
nucleon case to the $\Lambda$. The $\Lambda$ wave function in the
conventional SU(6) quark model is written as
\begin{equation}
|\Lambda^{\uparrow} \rangle =\frac{1}{2\sqrt{3}} [(u^{\uparrow} d
^{\downarrow} + d^{\downarrow} u ^{\uparrow}) -(u^{\downarrow} d
^{\uparrow} + d ^{\uparrow} u ^{\downarrow} )] s^{\uparrow} +
(\mathrm{cyclic ~~permutation}), \label{SU6}
\end{equation}
which can be reorganized into the SU(6) quark-diquark model wave
function,
\begin{eqnarray}
|\Lambda^{\uparrow} \rangle &=& \frac{1}{\sqrt{12}} [V_0(ds) u
^{\uparrow} - V_0(us) d ^{\uparrow} - \sqrt{2} V_{+1}(ds) u
^{\downarrow} + \sqrt{2} V_{+1}(us) d ^{\downarrow} \nonumber \\
&+& S(ds) u ^{\uparrow} + S(us) d ^{\uparrow} -2  S(ud) s
^{\uparrow}],
\label{SU6D}
\end{eqnarray}
where $V_{s_z}(q_1 q_2)$ stands for a $q_1 q_2$ vector diquark
Fock state with third spin component $s_z$, and $S(q_1 q_2)$
stands for a $q_1q_2$ scalar diquark Fock state.

 From Eq.~(\ref{SU6D}) we get the unpolarized quark distributions for the
three valence $u$, $d$, and $s$ quarks for the $\Lambda$,
\begin{equation}
\begin{array}{clcr}
&u_v(x)=d_v(x)=\frac{1}{4} a_{V(qs)}(x) +\frac{1}{12}
a_{S(qs)}(x);
\\
&s_v(x)=\frac{1}{3} a_{S(ud)}(x),
\end{array}
\end{equation}
where $a_{D(q_1 q_2)}(x) \propto  \int [\mathrm{d}^2
\vec{k}_\perp] |\varphi (x, \vec{k}_\perp)|^2$ ($D=S$ or $V$)
denotes the amplitude for the quark $q$ being scattered while the
spectator is in the diquark state $D$, and is normalized such that
$\int_0 ^1 a_{D(q_1 q_2)}(x) \mathrm{d} x =3$. We assume the $u$
and $d$ symmetry $D(qs)=D(us)=D(ds)$, from the $u$ and $d$
symmetry inside $\Lambda$. Similarly, the quark spin distributions
for the three valence quarks can be expressed as,
\begin{equation}
\begin{array}{clcr}
&\Delta u_v(x)=\Delta d_v(x)=-\frac{1}{12} a_{V(qs)}(x)
+\frac{1}{12} a_{S(qs)}(x);
\\
&\Delta s_v(x)=\frac{1}{3} a_{S(ud)}(x).
\end{array}
\end{equation}
In order to perform the calculation, we employ the
Brodsky-Huang-Lepage prescription \cite{BHL} for the light-cone
momentum space wave function for the quark-spectator $\varphi (x,
\vec{k}_\perp) = A_D \mathrm{exp}\{-\frac{1}{8\alpha_D^2}
[\frac{m_q^2+\vec{k}_\perp ^2}{x} +
\frac{m_D^2+\vec{k}_\perp^2}{1-x}]\}$, with parameters (in units
of MeV) $m_q=330$ for $q=u$ and $d$, $m_s=480$, $\alpha_D=330$,
$m_{S(ud)}=600$, $m_{S(qs)}=750$, and $m_{V(qs)}=950$, following
Ref.\cite{Ma96}. The differences in the diquark masses
$m_{S(ud)}$, $m_{S(qs)}$, and $m_{V(qs)}$ cause the symmetry
breaking between $a_{D(q_1 q_2)}(x)$ in a way that $a_{S(ud)}(x) >
a_{S(qs)}(x) > a_{V(qs)}(x)$ at large $x$.

Thus the quark-diquark model predicts, in the limit $x \to 1$,
that $u(x)/s(x) \to 0$ for the unpolarized quark distributions,
$\Delta s(x)/s(x) \to 1$ for the dominant valence $s$ quark which
provides the quantum numbers of strangeness and spin of the
$\Lambda$ , and also $\Delta u(x)/u(x) \to 1$ for the non-dominant
valence $u$ and $d$ quarks.

We now look at the pQCD based analysis of the quark distributions
from minimally connected tree graphs of hard gluon exchanges
\cite{Far75,Bro95}. In
the region $x \to 1$ such approach can give rigorous predictions for the
behavior of distribution functions \cite{Bro95}. In particular, it
predicts ``helicity retention'', which means that the helicity of
a valence quark will match that of the parent nucleon. 
Explicitly, the quark distributions of a hadron $h$ have been 
shown to satisfy the counting rule \cite{countingr},
\begin{equation}
q_h(x) \sim (1-x)^p, \label{pl}
\end{equation}
where
\begin{equation}
p=2 n-1 +2 \Delta S_z.
\end{equation}
Here $n$ is the minimal number of the spectator quarks, and
$\Delta S_z=|S_z^q-S_z^h|=0$ or $1$ for parallel or anti-parallel
quark and hadron helicities, respectively \cite{Bro95}. 
Therefore the anti-parallel helicity quark distributions are 
suppressed by a relative factor $(1-x)^2$, and $\delta 
q(x)/q(x) \to 1$ as $x \to 1$. A further input into the 
model, explained in detail in Ref.~\cite{Bro95}, is to retain 
the SU(6) ratios only for the parallel helicity distributions at 
large $x$, since in this region SU(6) is broken into 
SU(3)$^{\uparrow}\times$SU(3)$^{\downarrow}$. With such power-law
behaviors of quark distributions, the ratio $d(x)/u(x)$ of the 
nucleon was predicted \cite{Far75} to be 1/5 as $x \to 1$, and 
this gives $F_2^n(x)/F_2^p(x)=3/7$, which is (comparatively) 
close to the quark-diquark model prediction $1/4$. From the 
different power-law behaviors for parallel and anti-parallel 
quarks, one easily finds that $\Delta q/q =1$ as $x \to 1$ for 
any quark with flavor $q$ unless the $q$ quark is completely 
negatively polarized \cite{Bro95}. Such prediction are quite 
different from the quark-diquark model prediction that $\Delta 
d(x)/d(x)=-1/3$ as $x \to 1$ for the nucleon \cite{Ma96,Mel96}. 
The most recent analysis \cite{Yang99} of experimental data for 
several processes supports the pQCD based prediction of the 
unpolarized quark behaviors $d(x)/u(x)=1/5$ as $x \to 1$, but 
there is still no definite test of the polarized quark behaviors 
$\Delta d(x)/d(x)$ since the $d$ quark is the non-dominant quark 
for the proton and does not play a dominant role at large $x$.
Furthermore, this pQCD based model has been successfully used in 
order to explain the large single-spin asymmetries found in many 
semi-inclusive hadron-hadron reactions, while other models have 
not been able to fit the data \cite{Bog99}.

We extend the pQCD based analysis from the proton case to the $\Lambda$.
 From the SU(6) wave function of the $\Lambda$ we get the explicit
total spin distributions for each valence quark,
\begin{equation}
\begin{array}{cllr}
&u^{\uparrow}=d^{\uparrow}=\frac{1}{2}; ~~~~
&u^{\downarrow}=d^{\downarrow}=\frac{1}{2};
\\
&s^{\uparrow}=1; &s^{\downarrow}=0,
\end{array}
\label{SU6q}
\end{equation}
for all values of $x$. In the pQCD based analysis at large $x$, 
the anti-parallel helicity distributions
are suppressed relative to the parallel ones, thus SU(6) is
broken to SU(3)$^{\uparrow}\times$SU(3)$^{\downarrow}$.
Also the relativistic effect due to the Melosh-Wigner rotation
causes a suppression in the helicity distributions observed
in deep inelastic scattering compared to the quark spin distributions
in the quark model \cite{MW}.
Nevertheless, our model still retains the ratio 
$u^{\uparrow}/s^{\uparrow}= 1/2$ at large $x$ 
\cite{Bro95}. Thus helicity retention plus broken SU(6) imply
immediately that $u(x)/s(x) \to 1/2$ and $\Delta q(x)/q(x) \to 1$
(for $q=u$, $d$, and $s$) for $x \to 1$, and therefore the flavor
structure of the $\Lambda$ near $x=1$ is a region in which 
accurate tests of the pQCD based approach can be made.

 From the power-law behaviors of Eq.~(\ref{pl}), we write down a
simple model formula for the valence quark distributions,
\begin{equation}
\begin{array}{clcr}
q^{\uparrow}(x)   \sim x^{-\alpha}(1-x)^3; ~~~~
q^{\downarrow}(x) \sim x^{-\alpha}(1-x)^5,
\end{array}
\label{pls}
\end{equation}
where $q^{\uparrow}(x)$ and $q^{\downarrow}(x)$ are the parallel
and anti-parallel quark helicity distributions and $\alpha$ is
controlled by Regge exchanges with $\alpha \approx 1/2$ for
nondiffractive valence quarks. This model is not meant to give a
detailed description of the quark distributions but to outline its
main features in the large $x$ region. Combining Eq.~(\ref{pls})
with Eq.~(\ref{SU6q}), we get,
\begin{equation}
\begin{array}{cllr}
&u^{\uparrow}(x)=d^{\uparrow}(x)=\frac{35}{64}x^{-\frac{1}{2}}(1-x)^3;
 ~~~~ &u^{\downarrow}(x)=d^{\downarrow}(x)=\frac{693}{1024}x^{-\frac{1}{2}}(1-x)^5;
\\
&s^{\uparrow}(x)=\frac{35}{32}x^{-\frac{1}{2}}(1-x)^3;
&s^{\downarrow}(x)=0,
\end{array}
\end{equation}
which obviously satisfies that $u(x)/s(x)=1/2$ and $\Delta
q(x)/q(x)=1$ (for $q=u$, $d$ and $s$) as $x \to 1$.

\begin{figure}[htb]
\begin{center}
\leavevmode {\epsfysize=4.5cm \epsffile{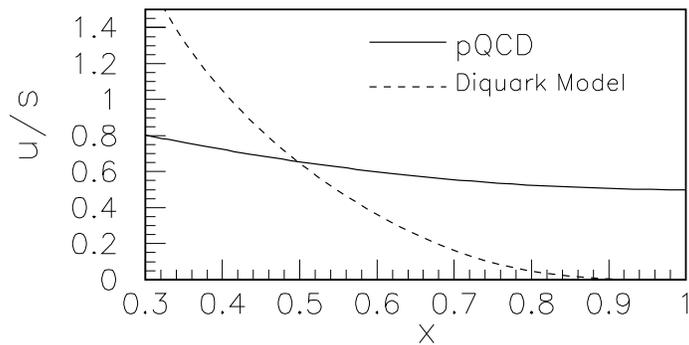}}
\end{center}
\caption[*]{\baselineskip 13pt The ratio $u(x)/s(x)$ of the
$\Lambda$ from the pQCD based approach (solid curve) and the SU(6)
quark-diquark
model (dashed curve). }\label{msy2f1}
\end{figure}

\begin{figure}[htb]
\begin{center}
\leavevmode {\epsfysize=4.5cm \epsffile{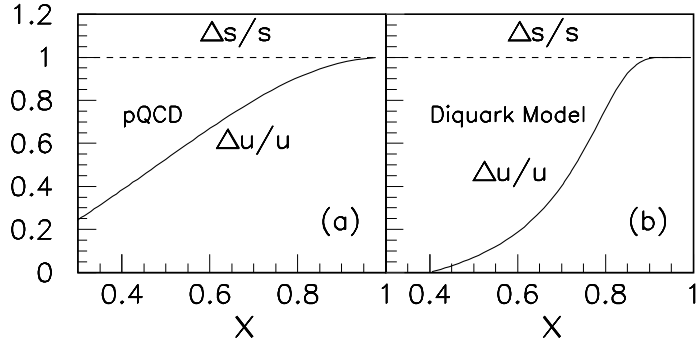}}
\end{center}
\caption[*]{\baselineskip 13pt The ratios $\Delta s(x)/s(x)$ for
the valence strange quark (dashed curves) and $\Delta u(x)/u(x)$
for the up and down valence quarks (solid curves) of the $\Lambda$
from (a) the pQCD based approach and from (b) the SU(6) quark-diquark
model. }
\label{msy2f2}
\end{figure}

In Fig.~\ref{msy2f1} we compare the $x$-dependence for the ratio
$u(x)/s(x)$ of the unpolarized quark distributions at large $x$ in
the two approaches and notice a significant difference between the
predictions from the quark-diquark model and the pQCD motivated
model. We also present in Fig.~\ref{msy2f2} the ratios $\Delta
s(x)/s(x)$ and $\Delta u(x)/u(x)$. We find that in both the
quark-diquark model and the pQCD motivated model, the dominant
valence $s$ quark is totally positively polarized, and the
non-dominant valence $u$ and $d$ quarks are also positively
polarized at large $x$, though they have net zero contributions to
the $\Lambda$ spin. This supports the viewpoint \cite{Jaf96} that
one cannot neglect the contribution from the $u$ and $d$ quarks,
in order to measure the strange quark polarization of a proton
$\Delta s(x)$ from the $\Lambda$ polarization of unpolarized
electron deep inelastic scattering (DIS) process on a polarized
proton target \cite{Lu95}. Though the qualitative features are
similar for the ratio $\Delta u(x)/u(x)$ in the two approaches,
the magnitude of $\Delta u(x)$ in the quark-diquark model should
be more suppressed than that from the pQCD based approach 
at large $x$ due to the
large suppression of $u(x)$. Therefore there is also difference in
the spin structure for different flavor valence quarks of the
$\Lambda$ near $x=1$.

We now discuss the processes that can serve to test the above
predictions. Direct measurement of the quark distributions of the
$\Lambda$ is difficulty, since the $\Lambda$ is a charge-neutral
particle which cannot be accelerated as incident beam and its
short life time makes it also difficult to be used as a target.
Thus one may extend the analysis of this work to a charged
hyperon, such as $\Sigma^{\pm}$ or $\Xi^-$, which might be used as
beam in Drell-Yan processes to test different predictions.
However, we know that the quark distributions inside a hadron are
related by crossing symmetry to the fragmentation functions of the
same flavor quark to the same hadron, by a simple reciprocity
relation \cite{GLR}
\begin{equation}
q_{h}(x) \propto D_q^h(z),
\label{GLR}
\end{equation}
where $z=2 p \cdot q/Q^2$ is the momentum fraction of the produced
hadron from the quark jet in the fragmentation process, and
$x=Q^2/2 p \cdot q$ is the Bjorken scaling variable corresponding
to the momentum fraction of the quark from the hadron in the DIS
process. Although such an approximate relation may be only valid
at a specific scale $Q^2$ near $x=1$ and $z=1$, 
it can provide a reasonable connection
between different physical quantities and lead to different
predictions about the fragmentations based on our understanding of
the quark structure of a hadron \cite{Ma99,Bro97}. 
 From another point of view, there are both experimental evidence and
theoretical
arguments to indicate the limitation of this relation 
for the physical application. Since our present
knowledge on the fragmentation functions is still poor, 
we may consider our study
as a phenomenological method to parameterize the quark
to $\Lambda$ fragmentation functions, and the validity
and reasonableness of the
method can be checked by comparison with the experimental
data on various quark to $\Lambda$ fragmentation functions.   
Thus we can use
various $\Lambda$ fragmentation processes to test different
predictions.

In principle we can test the different predictions by a
measurement of a complete set of quark to $\Lambda$ fragmentation
functions. One promising method to obtain a complete set of
polarized fragmentation functions for different quark flavors is
based on the measurement of the helicity asymmetry for
semi-inclusive production of $\Lambda$ hyperons in $e^+e^-$
annihilation on the $Z^0$ resonance \cite{Bur93}. There is also a
recent suggestion \cite{Ma99} to measure a complete set of quark
to $\Lambda$ unpolarized and polarized fragmentation functions for
different quark flavors by the systematic exploitation of
unpolarized and polarized $\Lambda$ and $\bar{\Lambda}$
productions in neutrino, antineutrino and polarized electron DIS
processes. However, in practice we do not need such systematic
studies of quark to $\Lambda$ fragmentations before we can test
the different predictions.

Some physical quantities related to the $\Lambda$ fragmentations
in specific regions can provide direct test of different
predictions. There have been suggestions that the ratio $\Delta
D_u^{\Lambda}(z) /D_u^{\Lambda}(z)$ can be measured from polarized
electron DIS process  \cite{Jaf96} and neutrino DIS process in the
region $y \simeq 1$ \cite{Ma99,Kot97}. Although the net $\Delta u$
might be zero or negative, and the magnitude also differs in
different predictions, the results of the present work tell us
that $\Delta D_u^{\Lambda}(z) /D_u^{\Lambda}(z)$ is positive at
large $z$, something unexpected from naive theoretical expectations.
There
have been also calculations \cite{Kot97,Bor98} of $\Lambda$
production in several processes, based on simple ansatz such as
$\Delta D_{q}^{\Lambda}(z) =C_{q}(z) D_{q}^{\Lambda}(z)$ with
constant coefficients $C_{q}$. From the present work we know that
there are many unreasonable assumptions on the detailed
$z$-dependence of the fragmentation in the previous calculations,
therefore their conclusions \cite{Kot97,Bor98} lack predictive
power and need to be re-checked with reasonable physical inputs
for various fragmentation functions.

The results presented in this letter should be considered as only
valid for the valence quarks, which are expected to play the
dominant role in the regions of $x \geq 0.4$ (or $z \geq 0.4$),
where we still need reliable data. In the small and medium $x$ (or
$z$) regions, the sea of the $\Lambda$ is expected to play an
important role, and there are many details that have to be
addressed in order to understand the physics in these regions. We
also notice that the study of this work can be directly extended
to other hadrons, such as $\Sigma$, $\Xi$, or even to heavier
flavor hadrons such as $\Lambda_c$, which may contribute as
backgrounds to the $\Lambda$ production or serve as new directions
to test different physics concerning hadron structure. We do not
expect to address all the vast issues in this brief letter, and 
more detailed analysis will be given elsewhere \cite{MSY3,MSY4}.

\begin{figure}[htb]
\begin{center}
\leavevmode {\epsfysize=6.5cm \epsffile{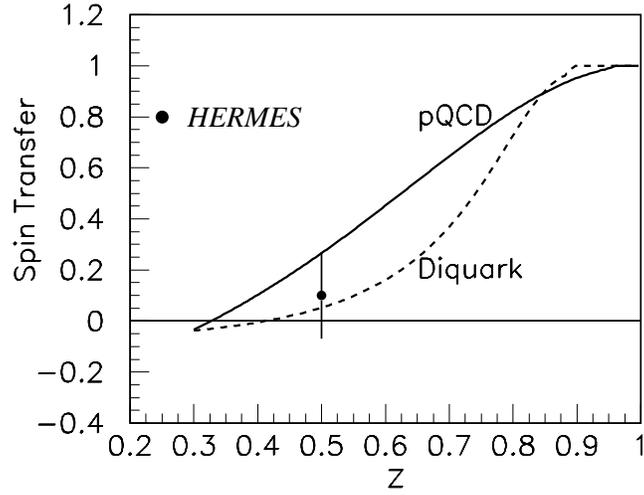}}
\end{center}
\caption[*]{\baselineskip 13pt 
The predictions of the longitudinal spin transfer
to the $\Lambda$ in deep inelastic scattering of polarized
lepton on the nucleon target from the simple pQCD based approach
(solid curve) and the SU(6) quark-diquark
model (dashed curve). The data point is the most recent
result by the HERMES collaboration \cite{HERMES}. }\label{msy2f3}
\end{figure}

We point out here that our prediction of positively polarized
$u$ and $d$ quarks inside $\Lambda$ at large $x$ has been supported
by the available data of $\Lambda$-polarization in $e^+e^-$-annihilation 
near the $Z$-pole \cite{MSY3} and also by the most recent HERMES
experiment on longitudinal spin transfer to the $\Lambda$ 
in deep elastic scattering of polarized positron on 
the nucleon target \cite{HERMES}. 
 From the results given in Ref.~\cite{MSY3}, we found that the
quark-diquark model gives a very good description of the available
experimental data of the $\Lambda$-polarization in
$e^+e^-$-annihilation near the $Z$-pole. The pQCD based analysis can 
also describe the data well by taking into account the suppression
due to the Melosh-Wigner rotation effect in the quark helicities
compared to the naive SU(6) quark model
spin distributions \cite{MW}. We also calculate and present here the
longitudinal 
spin transfer to the $\Lambda$ in deep elastic scattering
of polarized lepton on the nucleon target defined by
\cite{Jaf96,HERMES}
\begin{equation}
D^{\Lambda}_{L L'}=\frac{\sum_q e^2_q q_N(x)\Delta D_q^{\Lambda}(z)}
{\sum_q e^2_q q_N(x)D_q^{\Lambda}(z)},
\end{equation}
where $e_q$ is the charge of the quark, $q_N(x)$ is the 
quark distribution inside the nucleon, and $D_q^{\Lambda}(z)$
and $\Delta D_q^{\Lambda}(z)$ are the unpolarized and polarized
quark to $\Lambda$ fragmentation functions calculated by 
the relation (\ref{GLR}). Our predictions are shown
in Fig.~\ref{msy2f3}, in comparison with the recent HERMES data
\cite{HERMES}. 
We notice that the available HERMES data point is consistent with both
the quark-diquark model and the pQCD based predictions within 
the present err-bar,
and this seems to
support the positive $u$ and $d$ polarizations at large $x$ inside
the $\Lambda$ predicted in our work. 
Unfortunately, it is still not possible to 
make a clear distinction between the two different predictions of 
the flavor and spin structure of the $\Lambda$ by the available
data with the present statistical precision, and also in each
model there is still some freedom to adjust the parameters for a
better fit of the detailed features \cite{MSY3}. It is
also necessary 
to point out that the available HERMES date should not be considered as
in contradiction with the prediction of nagative polarizations of
$u$ and $d$ quarks inside the $\Lambda$ \cite{Bur93}, since
the positive polarizations of $u$ and $d$ at large $x$ do not rule out
the net negative $u$ and $d$ polarizations inside the $\Lambda$ 
integrated over whole $x=0 \to 1$.

In summary, we studied the flavor and spin structure of the
$\Lambda$ at large $x$ and found it is a region that can provide
clean tests of different predictions. We also found that the
up and down quarks should be positively polarized at large $x$, 
even though their net spin contributions
to the $\Lambda$ might be zero or negative. 
This prediction has been supported by the available data
of $\Lambda$-polarization in $Z$ decays and also
by the most recent HERMES result of spin transfer to the $\Lambda$
in deep elastic scattering of polarized lepton on the nucleon target.
Thus $\Lambda$ Physics is
a frontier with rich physics content that deserves attention both
theoretically and experimentally.

{\bf Acknowledgments: }
This work is partially supported by
National Natural Science Foundation of China under Grant
No.~19605006, No.~19975052, and No.~19875024, and by Fondecyt (Chile)
postdoctoral fellowship 3990048, by Fondecyt (Chile) grant 1990806
and by a C\'atedra Presidencial (Chile).

\end{document}